\documentclass[5p]{elsarticle}

\usepackage{graphicx,color}
\usepackage[hidelinks]{hyperref}
\usepackage{bm}
\usepackage{amsmath}
\usepackage{amssymb}
\usepackage[sort&compress, capitalise]{cleveref}
\usepackage{booktabs}
\usepackage{float}

\newcommand{\GeV}{$\,\text{GeV}$}
\newcommand{\itk}{\mathit k}
\newcommand{\wk}{\mathit{w}_\itk}
\newcommand{\bfa}{\boldsymbol{a}}

\usepackage[normalem]{ulem}

\newcommand*\oline[1]{%
   \vbox{%
     \hrule height 0.5pt%                  % Line above with certain width
     \kern0.4ex%                          % Distance between line and content
     \hbox{%
       \kern-0.15em%                        % Distance between content and left side of box, negative values for lines shorter than content
       \ifmmode#1\else\ensuremath{#1}\fi%  % The content, typeset in dependence of mode
       \kern-0.15em%                        % Distance between content and left side of box, negative values for lines shorter than content
     }% end of hbox
   }% end of vbox
}

\begin{document}
\title{Role of the Soffer bound in determination of transversity and the tensor charge}
%\title{A study on transversity, the Soffer bound, and \new{the} tensor charge of the nucleon}

% \title{Transversity function from SIDIS data and role of the Soffer bound}

\date{\today}

\author[add1,add2]{Umberto D'Alesio}
\ead{umberto.dalesio@ca.infn.it}

\author[add1,add2,add3]{Carlo Flore\corref{cor1}}
\address[add1]{Dipartimento di Fisica, Universit\`a di Cagliari, Cittadella Universitaria, I-09042 Monserrato (CA), Italy}
\address[add2]{INFN, Sezione di Cagliari, Cittadella Universitaria, I-09042 Monserrato (CA), Italy}
\address[add3]{Laboratoire de Physique des 2 Infinis Irène Joliot-Curie (IJCLab), CNRS, Université Paris-Saclay, 91406 Orsay Cedex, France}
\ead{flore@ipno.in2p3.fr}
\cortext[cor1]{Corresponding author}

\author[add4,add5]{Alexei Prokudin}
\address[add4]{Division of Science, Penn State University Berks, Reading, Pennsylvania 19610, USA}
\address[add5]{Theory Center, Jefferson Lab, 12000 Jefferson Avenue, Newport News, Virginia 23606, USA}
\ead{prokudin@jlab.org}

\begin{abstract}
The transversity and the tensor charge of the nucleon, currently under active investigation experimentally and theoretically, are fundamental quantities in hadron physics as well as for our comprehension of the nucleon structure. Some tension between the values of the tensor charge, as computed on the basis of phenomenological extractions and lattice QCD calculations, has been observed. In this letter, by means of an explicit example, we study the role of assumptions, usually adopted in phenomenological parametrizations, and we show that, by relaxing some of them, such a tension could be eased.
\end{abstract}

\begin{keyword}
transversity function \sep Soffer bound \sep tensor charge \sep QCD \sep JLAB-THY-19-3130
\end{keyword}

\maketitle
% \tableofcontents

\section{\label{sec:intro}Introduction}
The collinear transversity function~\cite{Ralston:1979ys}, $h_1^q(x)$, together with the unpolarized $f_{q/p}(x)$ and the helicity $g_{1L}^q(x)$ parton distribution functions, describe the collinear structure of a spin-$1/2$ hadron at leading twist. Unlike $f_{q/p}(x)$ and $g_{1L}^q(x)$, $h_1^q(x)$, being a chiral-odd quantity, cannot be directly accessed in inclusive deep-inelastic scattering processes (DIS), as another chiral-odd function is needed to form a chiral-even observable. At present, transversity has been extracted~\cite{Anselmino:2007fs,Anselmino:2013vqa, Kang:2015msa,Lin:2017stx} in Semi Inclusive Deep Inelastic Scattering (SIDIS) processes in combination with the Collins fragmentation function (FF)~\cite{Collins:1992kk}, or in two-hadron production in combination with a polarized dihadron fragmentation function~\cite{Jaffe:1997hf,Radici:2001na,Bacchetta:2012ty,Radici:2018iag, Benel:2019mcq}.

The possibility of accessing transversity in double polarized Drell-Yan process and a careful study of its properties and related sum rules were explored by Jaffe and Ji in Ref.~\cite{Jaffe:1991kp}. The $Q^2$ evolution of transversity was investigated by Artru and Mekhi in Ref.~\cite{Artru:1989zv} in leading order (LO) QCD. Soffer derived a positivity bound for transversity~\cite{Soffer:1994ww}, referred to as Soffer bound (SB). 
In Ref.~\cite{Barone:1997fh} Barone showed that the SB, if true at some initial scale $Q_0$, is preserved by QCD evolution at LO. Then,
Vogelsang, in Ref.~\cite{Vogelsang:1997ak}, extended this result showing that SB is preserved at next-to-leading-order (NLO) accuracy as well.

The validity of the bound itself was questioned by Ralston in Ref.~\cite{Ralston:2008sm}. On the other hand, it is a suitable tool and different research groups have used the SB in their phenomenological extractions~\cite{Anselmino:2015sxa,Kang:2015msa,Radici:2018iag,Benel:2019mcq}. The $x$-dependent part of $h_1^q$ is usually parametrized in terms of such bound at $Q_0$, the initial scale of the analysis.  
By imposing suitable constraints on the free parameters, the Soffer bound is automatically fulfilled throughout the fitting procedure. At variance, the recent study of Ref.~\cite{Benel:2019mcq} adopts the method of Lagrange multipliers to constrain transversity with a flexible parametrization of $h_1^q$.

Studies of transversity and the tensor charge are important for Beyond Standard Model (BSM) searches. Indeed, the isovector tensor charge, $g_T$, is related to potential tensor interactions in the electroweak sector~\cite{Cirigliano:2009wk, Bhattacharya:2011qm, Courtoy:2015haa, Yamanaka:2017mef}, and it is usually calculated on the lattice, as a matrix element over the full $x$ range, or by integrating the extracted transversity functions from phenomenological analyses. With respect to lattice QCD estimates, the latest phenomenological extractions of the transversity function~\cite{Anselmino:2007fs, Anselmino:2013vqa, Anselmino:2015sxa, Bacchetta:2012ty, Radici:2015mwa, Radici:2018iag,Lin:2017stx} seem to show a tension \cite{Radici:2019myq} on the estimated values of $g_T$ as well as on individual contributions from up, $\delta u$, and down, $\delta d$, quarks (see \cref{eq:deltaq-def1} below).

To this extent, it is interesting to check what is the role of the underlying assumptions adopted in phenomenological analyses. In this letter, by using an explicit example, we explore the impact of loosening some of these choices. This would bring us to analyse several aspects, such as the parameter-space exploration, whether we observe the violation of the Soffer bound in existing data and how all of this traduces into (isovector) tensor charge estimates.

The rest of the letter is organized as follows: in Section~\ref{sec:fit} we present the results of global fits of the transversity function from SIDIS and $e^+e^-$ data, obtained in the framework of the transverse momentum dependent (TMD) approach. Then, in Section~\ref{sec:tensor-charge}, we will investigate the impact of these results in estimating the tensor charges. Conclusions and comments are finally gathered in Section~\ref{sec:conclusions}.

\section{\label{sec:fit} Transversity from SIDIS data and role of the Soffer bound}

The bound~\cite{Soffer:1994ww}, derived by Soffer, reads:
\begin{equation}\label{eq:SB}
 |h_1^q(x, Q^2)| \leq \frac{1}{2} \left[f_{q/p}(x, Q^2) + g_{1L}^q(x, Q^2)\right] \equiv {\rm SB}(x, Q^2),
\end{equation}
where $f_{q/p}(x)$ and $g_{1L}^q(x)$ are respectively the unpolarized and the helicity parton distribution functions (PDFs).

Transversity has been extracted~\cite{Airapetian:2009ae,Alekseev:2008aa,Adolph:2014zba,Qian:2011py} in SIDIS, by analysing the so-called Collins asymmetry:
\begin{equation}
    A_{UT}^{\sin( \phi_h+\phi_S)} = \frac{2 (1-y)}{1+(1-y)^2}\frac{F_{UT}^{\sin(\phi_h+\phi_S)}}{F_{UU}} \; ,
\end{equation}
where $y$ is the fractional energy loss of the incident lepton, $F_{UU}={\cal{C}}[f_1D_1]$ is the unpolarized structure function and $F_{UT}^{\sin(\phi_h+\phi_S)}={\cal C}[h_1 H_1^\perp]$~\cite{Kotzinian:1994dv,Mulders:1995dh,Bacchetta:2006tn} is the polarized structure function of the SIDIS cross section, given as a convolution (over the unobserved transverse momenta) of the TMD transversity distribution, $h_1^q$, and the Collins FF, $H_1^\perp$. 

In order to unravel the transversity, one has to gather additional information on the Collins FFs, that could be accessed in $e^+e^-\to h_1 h_2 X$ processes via a $\cos (2 \phi_0)$ modulation, $A_0^{UL(C)}\propto {\cal C}[\bar H_1^\perp H_1^\perp]$~\cite{Boer:1997mf}. This was measured at the energy $\sqrt{s} \simeq 10.6$ GeV by the BELLE \cite{Seidl:2008xc} and the BABAR \cite{TheBABAR:2013yha} Collaborations as well as by the BESIII~\cite{Ablikim:2015pta} Collaboration, at a lower energy, $\sqrt{s} \simeq 3.65$ GeV.

Collins asymmetries in SIDIS and $e^+e^-$ processes at low values of transverse momentum (of the final hadron or of the almost back-to-back hadron pair) are formally expressed in terms of a TMD factorization approach~\cite{Bacchetta:2006tn,Boer:1997mf}. The TMD transversity function  $h_1^q(x,k_\perp)$, related to its collinear counterpart $h_1^q(x)$, was extracted in a series of global TMD fits of SIDIS and $e^+ e^- \to h_1 h_2 X$ data~\cite{Anselmino:2007fs, Anselmino:2013vqa, Anselmino:2015sxa}.

Complementary information on the collinear transversity function has been obtained also in the context of a collinear framework, for instance by considering its convolution with dihadron FFs in pion-pair production in SIDIS~\cite{Bacchetta:2012ty, Radici:2015mwa, Benel:2019mcq} and in polarized $pp$ collisions~\cite{Radici:2018iag}.

\subsection{\label{subsec:h1-fit} Fitting the TMD transversity function}

In this Section we present the results of our fits performed within a TMD approach. We will discuss and quantify the influence of initial assumptions and their impact on the extracted transversity functions.

Our analysis has been carried out following the approach of Ref.~\cite{Anselmino:2015sxa}, to which we refer the reader for all explicit expressions of the observables within the adopted parametrization. This somehow conservative choice will allow for a direct comparison with the results of  Ref.~\cite{Anselmino:2015sxa}, and will also help us to assess the role of assumptions for the fit parameters. Here we will highlight the differences with respect to the analysis of Ref.~\cite{Anselmino:2015sxa} starting from the new dataset: in addition to the SIDIS $A_{UT}^{\sin(\phi_h + \phi_S)}$ data from HERMES off a proton target~\cite{Airapetian:2010ds} and COMPASS off proton~\cite{Adolph:2014zba} and deuteron~\cite{Alekseev:2008aa} targets, and $e^+ e^-$ $A_0^{UL(C)}$ data from Belle~\cite{Seidl:2008xc} and Babar~\cite{TheBABAR:2013yha}, we have also included the latest BESIII data~\cite{Ablikim:2015pta} for $e^+ e^-$ azimuthal correlations. This results in a total number of datapoints $N_{\rm{pts}} = 278$.

In the analysis of Ref.~\cite{Anselmino:2015sxa} and Refs.~\cite{Anselmino:2007fs, Anselmino:2013vqa}, transversity was parametrized a\-dopt\-ing the usual Gaussian ansatz, with factorized $x$ and $k_\perp$ dependences, as
\begin{equation}\label{eq:h1(x)-SB-Q0}
 h_1^q(x, k^2_\perp) = h_1^q (x) \frac{e^{-k^2_\perp / \langle k^2_\perp \rangle}}{\pi \langle k^2_\perp\rangle}.
\end{equation}
We will use $\langle k^2_\perp \rangle = 0.57$ GeV$^2$, as extracted for the unpolarized TMD distributions from HERMES multiplicities in Ref.~\cite{Anselmino:2013lza}. The $x$-dependent part of transversity is usually parametrized~\cite{Anselmino:2007fs, Anselmino:2013vqa, Anselmino:2015sxa} at the initial scale $Q_0^2$ in terms of the Soffer bound, as
\begin{equation}\label{eq:h1(x)-SB}
 h_1^q(x, Q_0^2) = \mathcal{N}^T_q(x)\, {\rm SB}(x,Q_0^2).
\end{equation}
For the Soffer bound, Eq.~\eqref{eq:SB}, we adopt one of the most recent extractions of the collinear helicity distributions, name\-ly the NLO DSSV set of Ref.~\cite{deFlorian:2009vb}. For consistency, for the collinear unpolarized PDFs and FFs we adopt the NLO CTEQ66 PDFs set~\cite{Nadolsky:2008zw} and the NLO DSS $2014$ pion FFs set~\cite{deFlorian:2014xna}.
A transversity DGLAP kernel is then employed to carry out the evolution up to higher values of $Q^2$, by using an appropriately modified version~\cite{Courtoy:2012ry,Prokudin:hoppet} of {\tt  HOPPET} code~\cite{Salam:2008qg}. We adopt $Q_0^2 = 1.69\,\text{GeV}^2$ as the input scale, with $\alpha_S(M_Z) \simeq 0.118$ according to the CTEQ66 scheme.
The ${\cal N}^{T}_q(x)$ factor in~\cref{eq:h1(x)-SB} is given by
\begin{equation}
 {\cal N}^{T}_q(x)=N^{T}_q x^{\alpha}(1-x)^\beta\,
\frac{(\alpha+\beta)^{\alpha+\beta}}{\alpha^\alpha \beta^\beta},
\quad (q = u_v,\,d_v)
\end{equation}
with the same $\alpha$ and $\beta$ parameters for the valence $u_v$ and $d_v$ transversity functions. 

Upon constraining 
\begin{equation}
|N^T_q| \leq 1\; ,
\label{eq:choice}
\end{equation}
the transversity functions automatically fulfill their corresponding Soffer bound in~\cref{eq:SB}. Such constraint, as shown below, plays an important role in the extraction of the transversity function. To study and quantify the influence of the choice in Eq.~\eqref{eq:choice} we will perform two fits of the data using (and not using) such a condition on $N^T_q$ parameters, i.e.~ensuring (not ensuring) the automatic fulfilment of the SB throughout the fit. In the following plots, we will respectively refer to these two cases as ``using SB'' or ``no SB''.

The Collins functions are parametrized as in Ref.~\cite{Anselmino:2015sxa}
\begin{equation}
\label{eq:Collins}
H_1^{\perp q}(z, p_\perp^2)  = {\cal N}^C_q (z) \frac{z m_h}{ M_C}\,\sqrt{2e}\,e^{-p_\perp^2/M_C^2}\,D_{h/q}(z, p_\perp^2)\,,
\end{equation}
with $q = \rm{fav}, \rm{unf}$ (favoured/unfavoured) and where $M_C$ is a free parameter with mass dimension. $D_{h/q}(z, p_\perp^2)$ is the unpolarized TMD fragmentation function
\begin{equation} 
 D_{h/q}(z, p_\perp^2) = D_{h/q}(z)\, \frac{e^{-p^2_\perp / \langle p^2_\perp \rangle}}{\pi \langle p^2_\perp\rangle}\; ,
\end{equation}
with $\langle p^2_\perp\rangle = 0.12$ GeV$^2$~\cite{Anselmino:2013lza}; for $D_{h/q}(z)$ we use the NLO DSS 2014 set~\cite{deFlorian:2014xna}. The ${\cal N}^C_q(z)$ factors are given by 
\begin{equation}
\begin{aligned}
 & {\cal N}^C_{\rm{fav}}(z) = N^C_{\rm{fav}}\, z^\gamma (1-z)^\delta \frac{(\gamma + \delta)^{\gamma + \delta}}{\gamma^\gamma \delta^\delta}\,, \\
 & \qquad\qquad{\cal N}^C_{\rm{unf}}(z) = N^C_{\rm{unf}}\,.
\end{aligned}
\end{equation}
For its importance and later use we also give the first moment of the Collins function~\cite{Meissner:2010cc} 
\begin{equation}\label{eq:H1perp-moment}
\begin{aligned}
    H_1^{\perp (1)\,q} (z) & = z^2 \int d^2 \bm{p}_\perp \frac{p_\perp^2}{2 m_h^2}\,H_1^{\perp q} (z, z^2 p_\perp^2) \\[1mm]
    & = \sqrt{\frac{e}{2}} \frac{1}{z m_h} \frac{M_C^3 \langle p^2_\perp \rangle}{\left(\langle p^2_\perp \rangle + M_C^2\right)^2}\,{\cal N}^C_q(z) D_{h/q}(z)\,,
    \end{aligned}
\end{equation}
where the last expression is obtained adopting the pa\-ram\-e\-tri\-za\-tion in Eq.~(\ref{eq:Collins}). 

In order to estimate the errors of the extracted functions, we will follow the procedure of Ref.~\cite{Sato:2016wqj}, and for a given observable $\cal{O}$ we compute the expectation value $\rm{E}[\cal{O}]$ and variance $\rm{V}[\cal{O}]$ as
\begin{equation}\label{eq:E}
 \rm{E}[\mathcal{O}] = \int \it{d^{n}\!a}\,\cal{P}(\bfa|\rm{data})\,\cal{O}(\bfa) \simeq \sum_{\mathit k} \wk\,\cal{O}(\bfa_{\mathit{k}})\,,
\end{equation}
\begin{equation}\label{eq:V}
\begin{aligned}
 \rm{V}[\cal{O}] & = \int \it{d^{n}\!a}\,\cal{P}(\bfa|\rm{data})\left(\cal{O}(\bfa) - \rm{E}[\cal{O}]\right)^2\\
 & \simeq \sum_{\itk}\wk\left(\cal{O}(\bfa_\itk) - \rm{E}[\cal{O}]\right)^2.
\end{aligned}
\end{equation}
$\cal{O}$ is a function of the $n$-dimensional parameter vector $\bfa$ with a multivariate probability density $\cal{P}(\bfa|\rm{data})$~\cite{Sato:2016wqj} for parameters $\bfa$ conditioned by existing experimental data. This can be written using Bayes' theorem as 
\begin{equation}
   \cal{P}(\bfa|\rm{data})=\frac{{\cal L} (\rm{data}| \bfa)\, \pi(\bfa)}{\mathit{Z}}\; ,
\end{equation}
where ${\cal L} (\rm{data}| \bfa)$ is the likelihood, $\pi(\bfa)$ is the prior, and $Z$ is the evidence.
We follow Refs.~\cite{Sato:2016wqj,Lin:2017stx} and discretize the integrals in Eqs.~(\ref{eq:E}) and~(\ref{eq:V}) that lead to the introduction of weights $\wk$ related to the corresponding $\chi^2$'s as
\begin{equation}\label{eq:weights}
 \wk = \frac{\rm{exp}\left[-\frac 12\,\chi^2(\bfa_\itk)\right] \pi(\bfa_\mathit{k})}{\sum\limits_k \wk}\; .
\end{equation}
The priors $\pi(\bfa)$ are obtained using the Monte Carlo (MC) procedure described in the Appendix A of Ref.~\cite{Anselmino:2008sga}, while the $N_{\rm set}$ parameter sets are generated by a multidimensional MC generator, utilizing the covariance matrix from the fit by MINUIT~\cite{James:1975dr}.

In particular, we start with $N_{\rm set}=10^5$ parameter sets $\bfa_k$ for the ``no SB" fit, so that $k\in[1,N_{\rm set}]$ in ~\cref{eq:E,eq:V}. We then implement the Soffer bound directly on the priors by keeping those sets with $|N_{u_v  (d_v)}^T| \le 1$: this results in $N_{\rm set}=15570$ for the ``using SB" fit. As we will discuss below, this procedure guarantees a proper exploration of the parameter space.  

\begin{figure}[t]
\centering
\includegraphics[width=.39\textwidth, bb = 16 12 341 238]{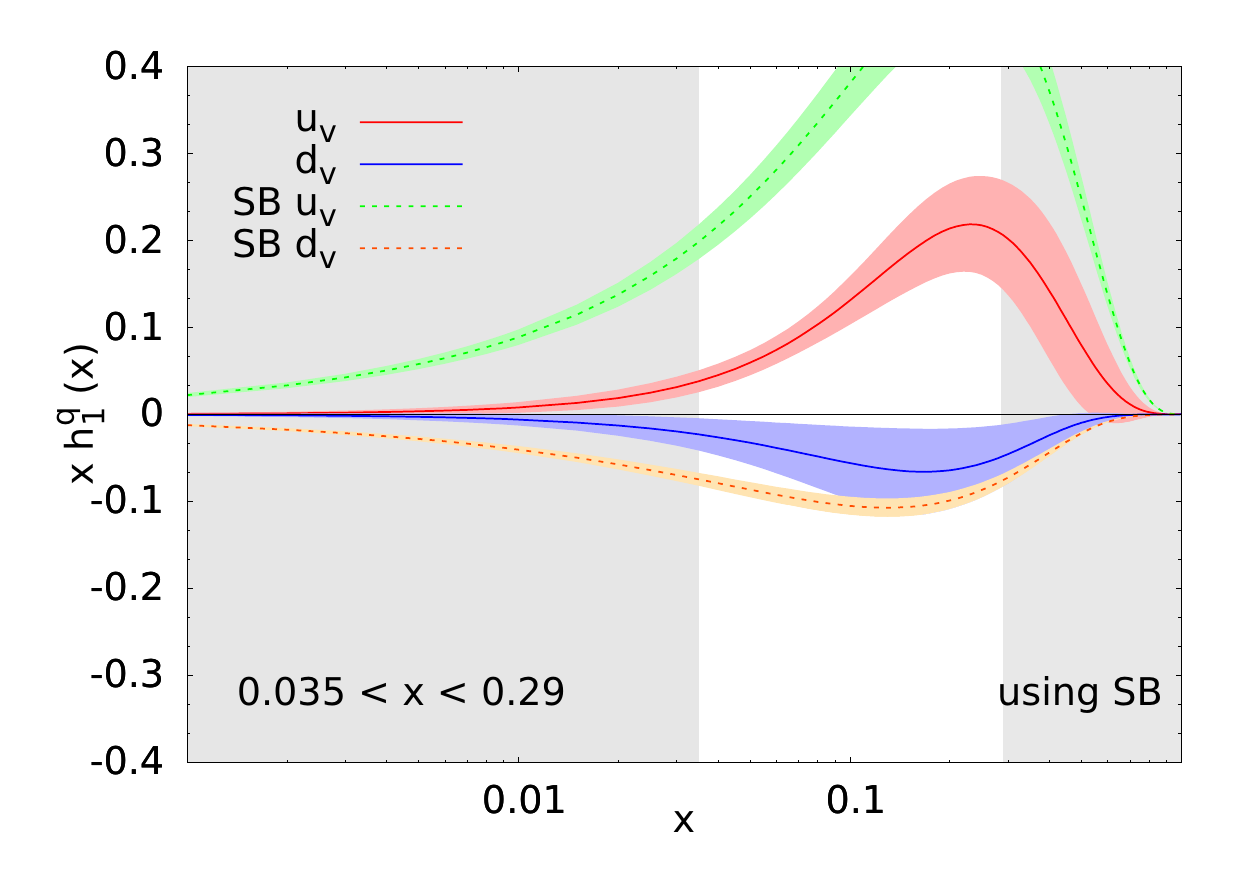}
\includegraphics[width=.39\textwidth, bb = 16 12 341 238]{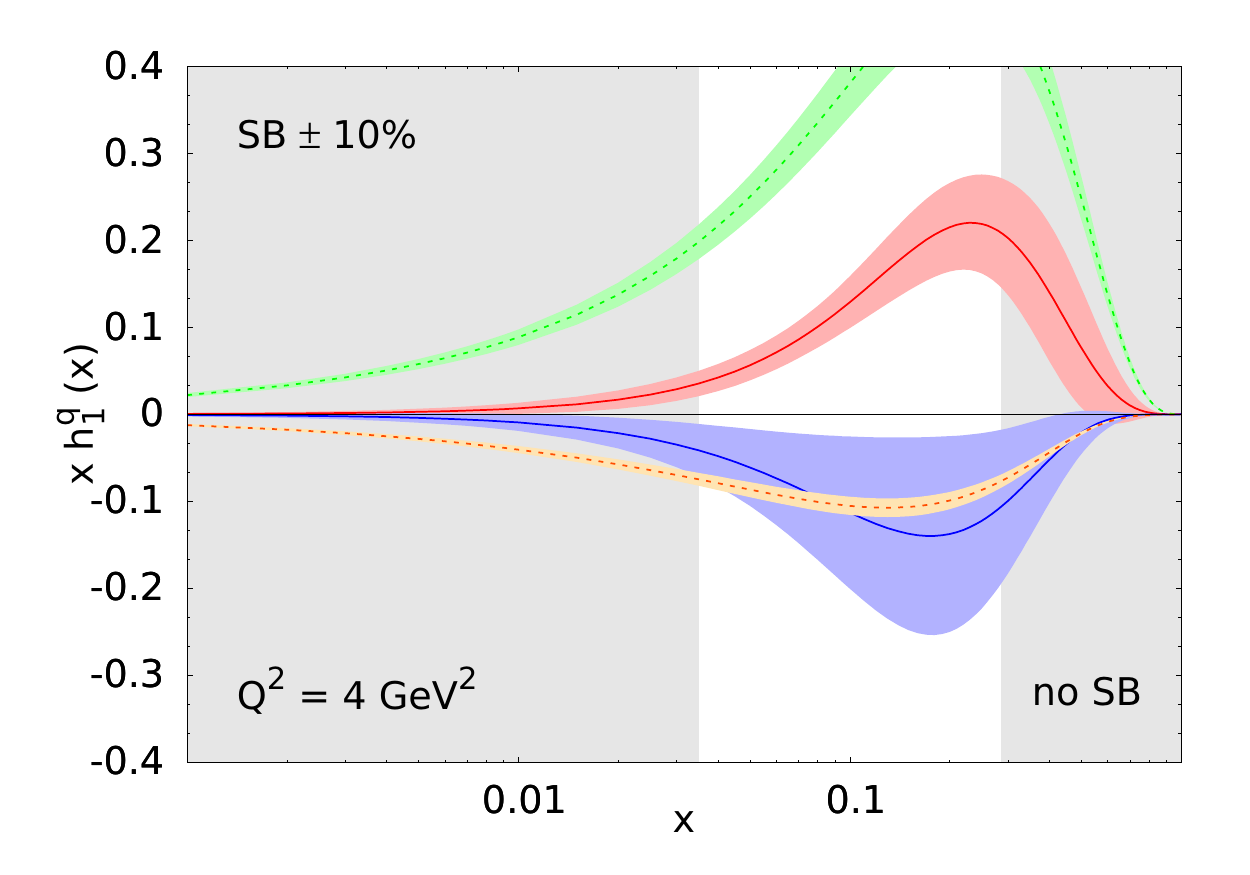}
\caption{Transversity functions for $u_v$ (red) and $d_v$ (blue) flavours from a global fit to SIDIS and $e^+ e^-$ data at $Q^2 = 4$ GeV$^2$. Upper panel: results with automatic fulfillment of the Soffer Bound ($|N^T_{u_v(d_v)}|\leq 1$). Lower panel: results with no constraint on $N^T_{u_v(d_v)}$. Error bands on the fitted functions are at $2\sigma$ level. The corresponding Soffer bound, computed with CTEQ66~\cite{Nadolsky:2008zw} PDFs and DSSV~\cite{deFlorian:2009vb} helicity distributions, is also shown for $u_v$ (green) and $d_v$ (orange), together with a $\pm 10\%$ variation. The white area represents the region of the bulk of the data; outside it no data are included in the fit.
}
\label{fig:transv-step1&3}
\end{figure}

The bands and the central lines in ~\cref{fig:transv-step1&3} are computed according to~\cref{eq:E,eq:V}. \cref{fig:transv-step1&3} shows the results of the two types of fits. The extracted transversity functions for $u_v$ and $d_v$ flavours, together with their $2\sigma$  uncertainty bands, are plotted at $Q^2 = 4$~GeV$^2$. The Soffer bound at $Q^2 = 4$ GeV$^2$ for the two flavours is also shown, with a $\pm 10\%$ variation representing an uncertainty estimate on their central values. The grey areas are the ones outside the existing data, that lies in the range $0.035 \lesssim x \lesssim 0.29$, corresponding to the smallest and highest $x$ value respectively probed by the HERMES and COMPASS Collaborations.

\begin{figure}[t]
\centering
\includegraphics[width=.39\textwidth, bb = 1 6 341 243]{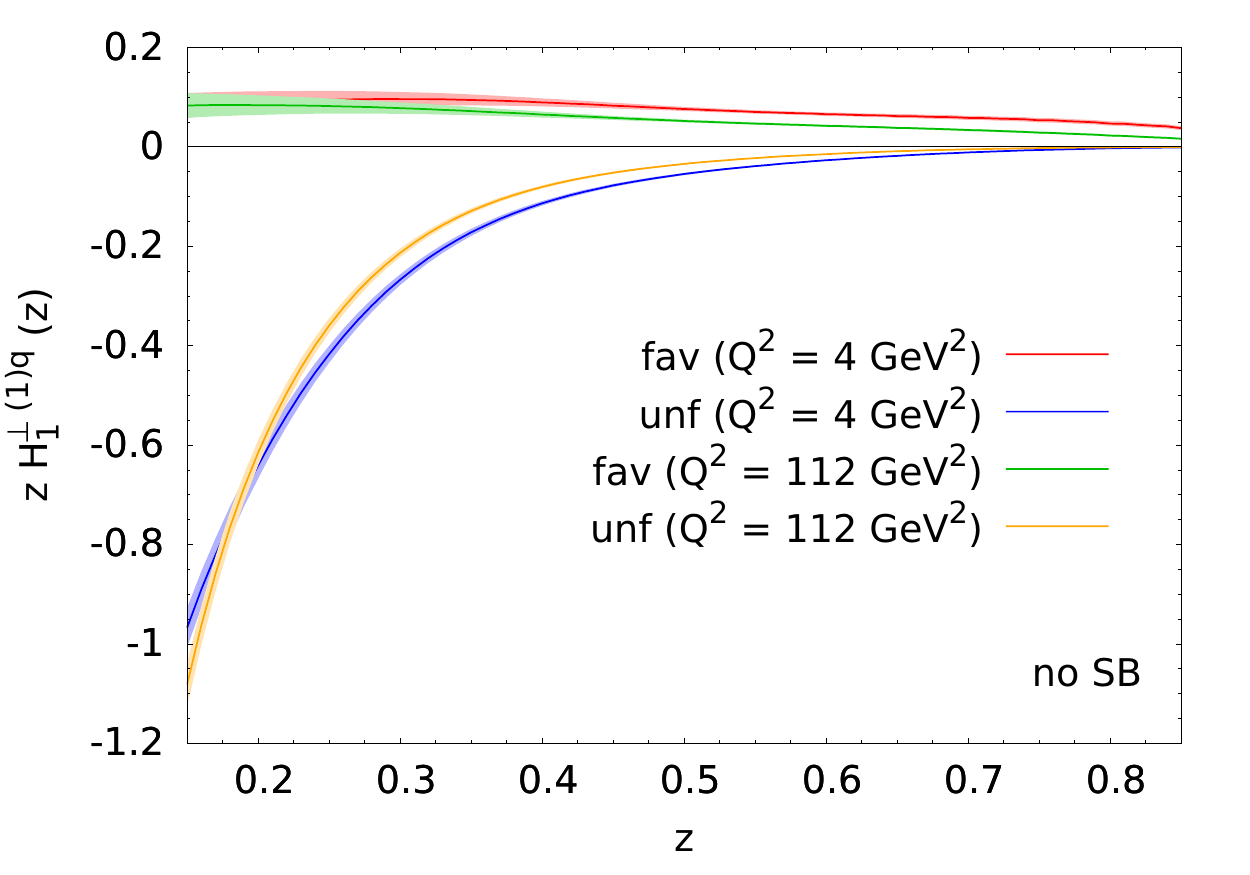}
\caption{First moments of the favoured and unfavoured Collins fragmentation functions from a global fit to SIDIS and $e^+ e^-$ data for the ``no SB" case. Curves at $Q^2 = 4$ GeV$^2$ (red and blue) and $Q^2 = 112$ GeV$^2$ (green and orange) are shown. Error bands on the fitted functions are at $2\sigma$ level. Notice that the results for the ``using SB" case are practically the same.}
\label{fig:Collins-step3}
\end{figure}

Besides the differences showed in~\cref{fig:transv-step1&3}, we underline that the two extractions have essentially the same statistical significance, rendering similar minimum $\chi^2$'s ($\chi^2_{\rm{min}} = 251.23$ and $\chi^2_{\rm{min}} = 250.21$ for the ``using SB" and ``no SB" cases, respectively), and essentially the same $\chi^2_{\rm{dof}}\simeq 0.93$ for the $N_{\rm{par}} = 9$ parameter fit. 
For completeness, in~\cref{fig:Collins-step3} we show the first moments of the Collins FFs as extracted without imposing the SB on the transversity functions. Notice that the corresponding ones obtained using the SB are practically the same. This means that the Collins FFs are strongly constrained by $e^+e^-$ data alone and are not so sensitive to the assumptions made on the transversity functions.

%%%%%%%%%%%%%%%%%%%%%%%%%%%%%%%%%%%%%%%%
%%%%%%%%%%%%%%%%%%%%%%%%%%%%%%%%%%%%%%%%
%%%%%%%%%%%%%%%%%%%%%%%%%%%%%%%%%%%%%%%%
%%%%%%%%%%%%%%%%%%%%%%%%%%%%%%%%%%%%%%%%

We start noticing that, since the helicity distribution for the $d_v$ quark flavour is negative, the corresponding SB is much more stringent with respect to the $u_v$ one. So, in extracting $h_1^{d_v}(x)$, there is less room for the parameters to vary. We also mention that, in all previous fits, $N^T_{d_v}$ was almost always saturating its lower bound~\cite{Anselmino:2013vqa, Anselmino:2015sxa}.

In the upper panel of~\cref{fig:transv-step1&3}, as expected, we observe two functions comparable to the existing extractions in Refs.~\cite{Anselmino:2007fs, Anselmino:2013vqa, Anselmino:2015sxa} and respecting the SB for both flavours used in the fit. For the extraction corresponding to ``no SB", lower panel of~\cref{fig:transv-step1&3}, we can note the following:
\begin{enumerate}[(i)]
 \item when relaxing the constraint on $N_T^{u_v}$, the corresponding transversity function does not essentially change with respect to the one in the upper panel;
 \item conversely, the $d_v$ transversity function tends to violate its Soffer bound, especially in the region where data are present;
 \item while the uncertainty bands of the two extracted $h_1^{u_v}$ are quite similar, there is a significant difference between the uncertainties of $h_1^{d_v}$.
\end{enumerate}

Now, we have to estimate the statistical significance of the violation of the SB for the down-quark transversity function shown in~\cref{fig:transv-step1&3}. To this aim, we use a simple $z$-score method to measure whether we are observing a statistically significant deviation from the zero hypothesis, i.e.~the fulfillment of the SB by the $d_v$ transversity function. The $z$-score is generally defined as
\begin{equation}
 z = \frac{x-\mu}{\sigma}\,,
\end{equation}
and tells us how many standard deviations $\sigma$ we are far from the mean $\mu$ for the point $x$. In our case, $x$ is the fitted value of the function, $\mu$ the corresponding SB value and $\sigma$ the estimated uncertainty on the function. For the extracted $d_v$ transversity function, we have $-0.9 \leq z \leq -0.3$ for the whole region, that means that the SB for down quark is well within 1$\sigma$ deviation and we can conclude that the violation is not statistically significant.

Another aspect is related to the exploration of the parameter space. While the uncertainty on the $u_v$ transversity function is essentially unchanged when relaxing the initial constraint, this is not the case for the $d_v$ one. 

\begin{figure}[t]
\centering
\includegraphics[width=.39\textwidth, bb = 9 14 341 236]{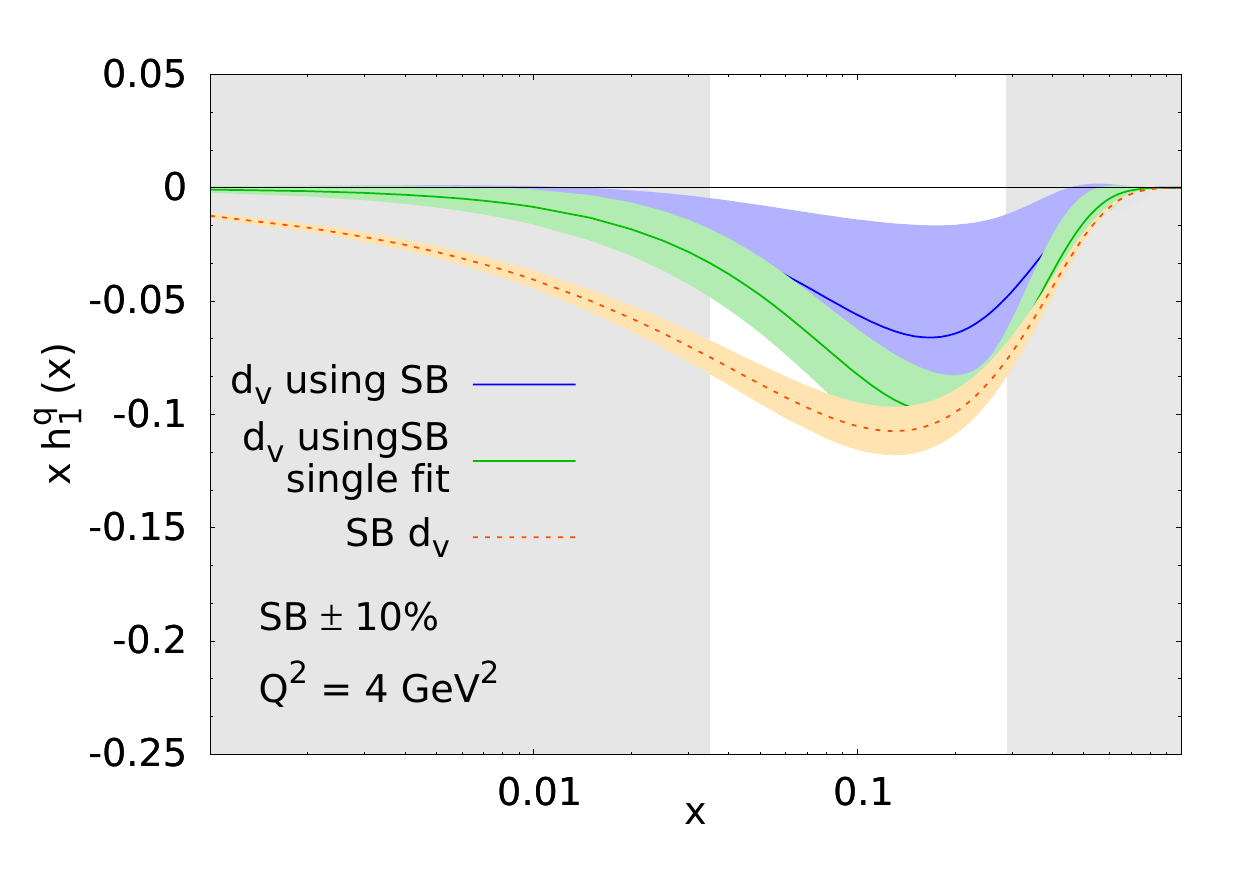}
\caption{Transversity functions for $d_v$ flavour from a global fit to SIDIS and $e^+ e^-$ data at $Q^2 = 4$ GeV$^2$. Blue band corresponds to usage of priors from Fig.~\ref{fig:transv-step1&3}, while green band corresponds to priors generated from constrained fit $|N^T_{d_v}| \le 1$ directly. Error bands on the fitted functions are at $2\sigma$. The corresponding Soffer bound, computed with CTEQ66~\cite{Nadolsky:2008zw} PDFs and DSSV~\cite{deFlorian:2009vb} helicity distributions, is also shown, together with a $\pm 10\%$ variation. The white area represents the region of the bulk of the data; outside it no datapoints are present in the fit.
}
\label{fig:transv-step4}
\end{figure}

Furthermore, as mentioned in Ref.~\cite{Lin:2017stx}, the exploration of the  parameter space starting from a single fit may lead to incorrect estimates of both mean values and errors of observables and/or extracted parameters.
We can demonstrate it explicitly as follows. 
We consider the constrained ``using SB" fit that turns in the saturation of the normalization parameter $N^T_{d_v} = -1$. 
The main point is that, when requesting the parameter $N^T_{d_v}$ to be limited between $-1$ and $+1$ (actually between any $a,\,b$), the MINUIT minimizer maps this region onto the unbound region from $-\infty$ to $+ \infty$. Once the fitted parameter is close to its lower bound value ($-1$), the internal parameter of the minimizer goes therefore to $-\infty$.  
This prevents to explore all regions in the parameter space compatible with the theoretical expectations and with the calculated $2\sigma$ error without imposing any bound on parameters.
If we now generate the priors using the covariance matrix, the resulting distributions show artificially small errors for the $d$-quark transversity, and where this saturates, as happens at $x\sim 0.2$, the error becomes extremely small (see green band in Fig.~\ref{fig:transv-step4}).
%%%%%%%%%%%%%%%%%%%%%%%%%%%%%%%%%%%%%%%%
%%%%%%%%%%%%%%%%%%%%%%%%%%%%%%%%%%%%%%%%
This behaviour is typical for constrained fits, see Fig.~4 of Ref.~\cite{Bacchetta:2012ty} or Fig.~8 of Ref.~\cite{Radici:2015mwa}. SIDIS process is dominated by $u$-quark contribution and thus one expects the relative precision for $d$ quark to be worse with respect to the one reachable for $u$ quark. This is clearly not the case for the $d$-quark green band in Fig.~\ref{fig:transv-step4} when compared to the corresponding one for $u$ quark in Fig.~\ref{fig:transv-step1&3}.

Indeed, there exist configurations, compatible with the SB, that are not explored when the constraint on the $N^T_q$ parameters is imposed directly in the fit.
This issue was mitigated in Refs.~\cite{Anselmino:2013vqa, Anselmino:2015sxa} by generating several hundred thousands of parameter sets and in Ref.~\cite{Benel:2019mcq} by using the Lagrange multiplier method instead of imposing direct constraints on the parameters.

\section{\label{sec:tensor-charge}Tensor charges}

The contribution to the tensor charge of the nucleon from quark $q$ is the first Mellin moment of the non-singlet quark combination 
\begin{equation}\label{eq:deltaq-def1}
 \delta q = \int_0^1\,\left[h_1^{q}(x) - h_1^{\bar q}(x)\right]\,dx.
\end{equation}
The isovector combination, $g_T$, is of particular interest and can be calculated relatively easily on the lattice~\cite{Alexandrou:2016hiy}:
\begin{equation}\label{eq:gT-def}
 g_T = \delta u - \delta d.
\end{equation}

In our analysis we have $h_1^{\bar q}\equiv 0$ and thus we compute valence quark tensor charges as
\begin{equation}\label{eq:deltaq-def}
 \delta q_v = \int_0^1\, h_1^{q_v}(x)\,dx.
\end{equation}

It is also useful to mention that truncated charges can be built, upon integrating in~\cref{eq:deltaq-def} between the experimental minimum and maximum $x$ values, $x_{\rm min}$ and $x_{\rm max}$ respectively.

As $g_T$ is related to BSM effects~\cite{Cirigliano:2009wk, Bhattacharya:2011qm, Courtoy:2015haa, Yamanaka:2017mef}, a phenomenological extraction of the transversity functions can be used in principle to put a limit on the strength of this potential non-standard interactions. At the same time, this can be also reached by using the lattice QCD estimates of $g_T$. For a comprehensive review of lattice results see Ref.~\cite{Alexandrou:2016hiy} and references therein.

At this point it is important to recall that some tension have been observed~\cite{Radici:2019myq} between phe\-nom\-e\-no\-log\-i\-cal estimates and lattice QCD calculations of $g_T$ and individual quark contributions. In fact, $\delta u_v$ values from phenomenology seem to be incompatible with the lattice ones, and $g_T$ values calculated on the lattice are found to be higher than the ones from most phenomenological analyses~\cite{Radici:2018iag}. Lattice results are approximately in the range $0.9 \lesssim g_T \lesssim 1.1$, and with very tiny errors, for instance $0.926(32)$ from a recent study in Ref.~\cite{Alexandrou:2019brg}. It is then interesting to explore the impact of the results presented in~\cref{sec:fit} on the phenomenological estimates of the tensor charges.

By integrating the two couples of extracted transversity functions of~\cref{fig:transv-step1&3}, we calculate for every MC set the corresponding tensor charges, $\delta u_v$ and $\delta d_v$, and thus the corresponding isovector tensor charge, $g_T$. The corresponding central values and errors are again computed according to~\cref{eq:E,eq:V}.

\begin{figure}[t]
\centering
\includegraphics[width=.39\textwidth, keepaspectratio, bb = 21 0 390 255]{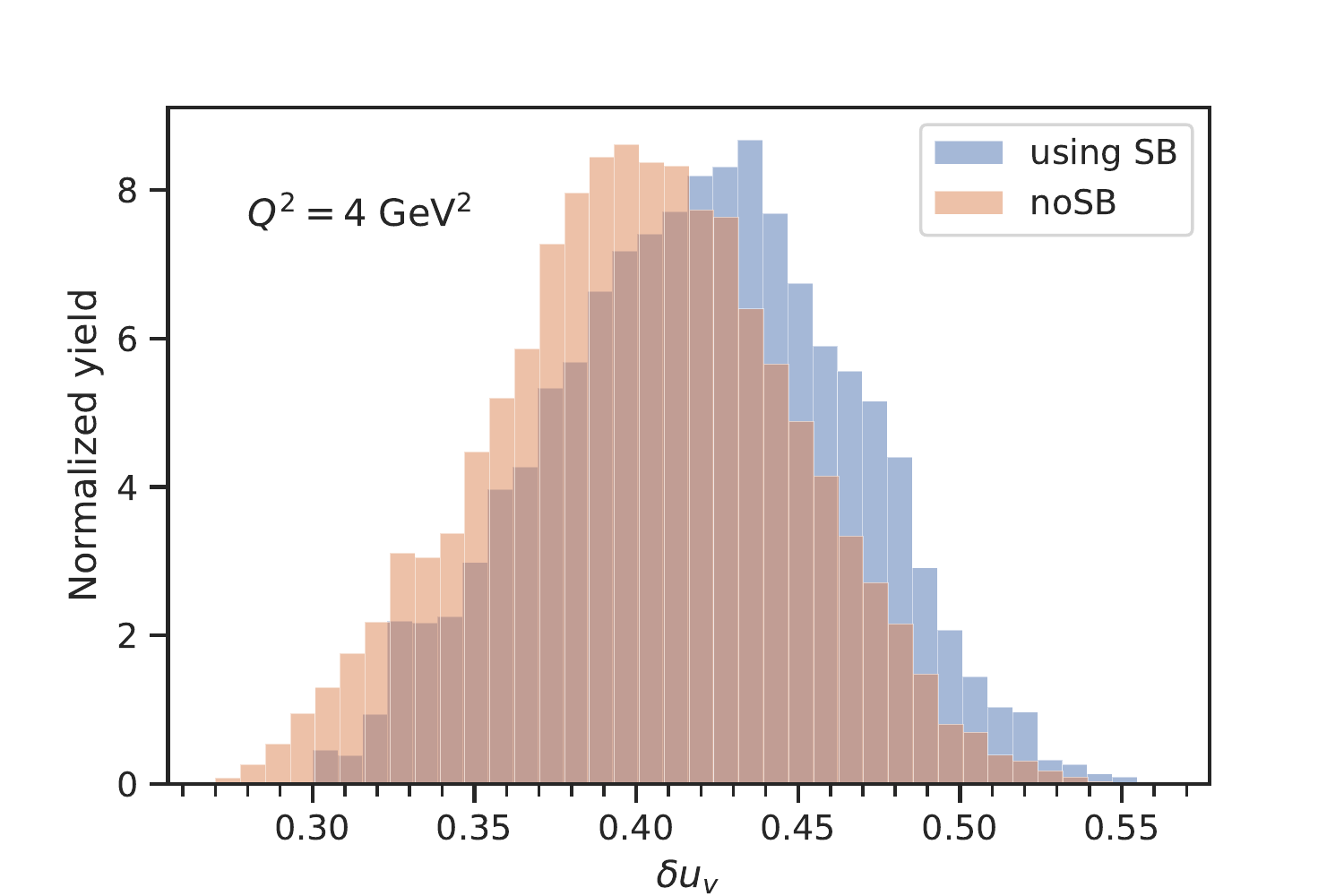}\\[6mm]
\includegraphics[width=.39\textwidth, keepaspectratio, bb = 21 1 390 255]{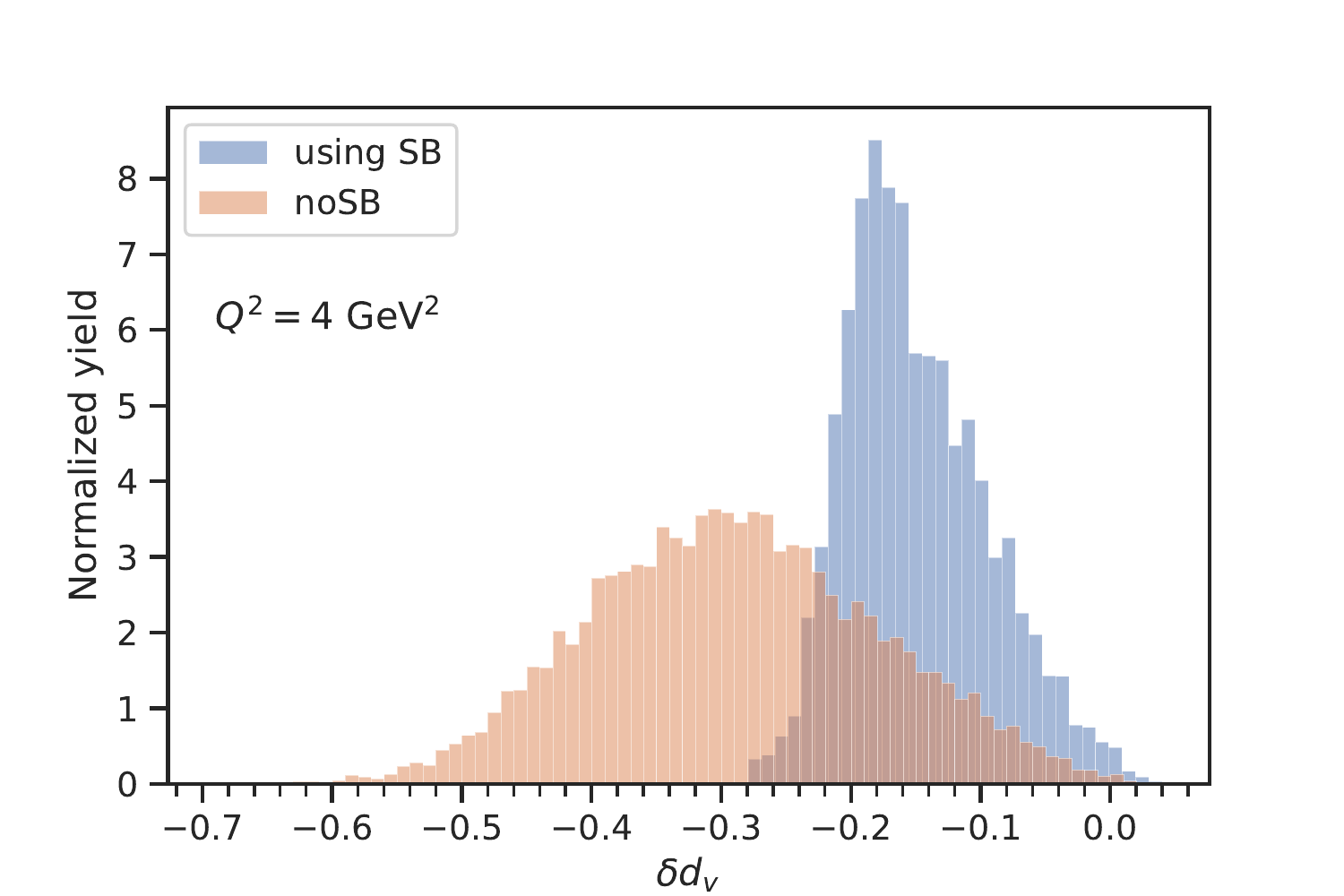} 
\caption{Distributions of the tensor charges for $u_v$ (upper panel) and $d_v$ (lower panel) at $Q^2 = 4\,\text{GeV}^2$. The tensor charges are calculated using the extracted transversity distributions of~\cref{fig:transv-step1&3}, integrated over the full range $0 \leq x \leq 1$. Labels ``using SB'' and ``no SB'' have the same meaning as in~\cref{fig:transv-step1&3}.}
\label{fig:deltaq-distr-step1&3}
\end{figure}

To begin with, we can check the effect of relaxing the hypothesis $|N^T_q| \leq 1$ on the tensor charge distributions. \cref{fig:deltaq-distr-step1&3} shows the distribution of $\delta u_v$ (upper panel) and $\delta d_v$ (lower panel) calculated at $Q^2 = 4\GeV^2$, the usual energy scale adopted to compare tensor charges calculated on the basis of phenomenological analyses and lattice QCD estimates. The labels ``using SB'' and ``no SB'' have the same meaning as in~\cref{fig:transv-step1&3}. As one could expect, when relaxing the initial constraint, the $\delta u_v$ distribution does not change much, thus reflecting the very small difference observed in the extracted $h_1^{u_v}$ in~\cref{fig:transv-step1&3}. At variance with this, the $\delta d_v$ distribution dramatically changes, reflecting once more what has been observed for the fitted $d_v$ transversity function in~\cref{fig:transv-step1&3}. 
%Again, by requesting the automatic fulfillment of the SB for the transversity function, we are preventing the minimizer to explore some configuration compatible to the $2\sigma$ error band. This translates into a very narrow and peaked distribution for the $\delta d_v$ ``using SB'' case. On the other hand, when looking at the ``no SB'' case, the $\delta d_v$ distribution is broader and flattened, reflecting the large uncertainty on the corresponding fitted transversity function.

For the individual quark distributions, we find that
both $\delta u_v$ and $\delta d_v$ are different from lattice computations, $0.716(28)$ and  $-0.210(11)$ respectively found in Ref.~\cite{Alexandrou:2019brg}, see Table~\ref{tab:results}.
Although these results do not ease the tension between phenomenological and lattice QCD estimates of $\delta u_v$ and $\delta d_v$, they actually have an effect on the isovector tensor charge estimates. 

\begin{figure}[t]
\centering
\includegraphics[width=.39\textwidth, keepaspectratio, bb = 21 1 390 255]{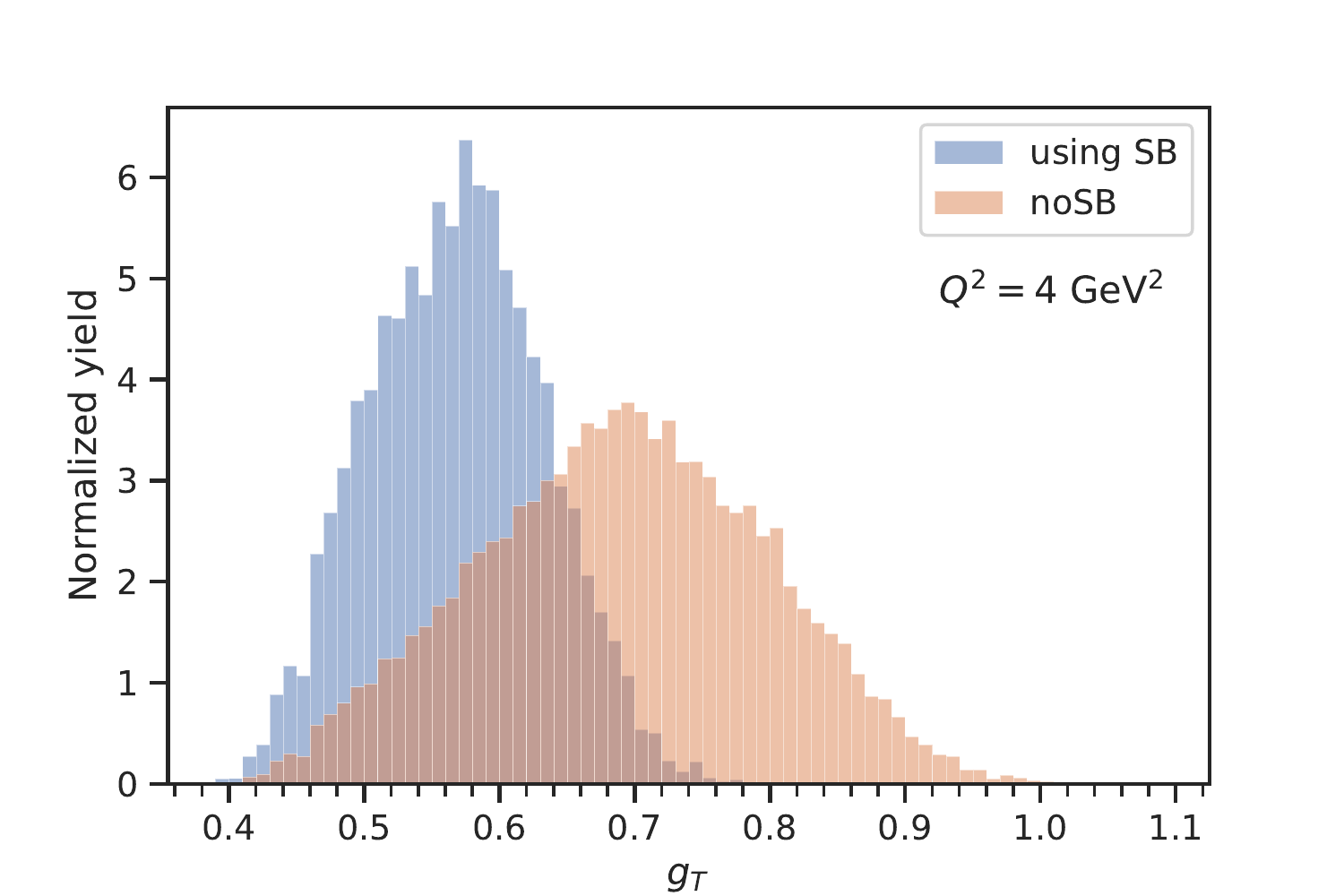}
\caption{Distributions of the isovector tensor charge, $g_T$, at $Q^2 = 4\GeV^2$. The calculation is performed using the extracted transversity distributions of~\cref{fig:transv-step1&3}, integrated over the full range $0 \leq x \leq 1$. Labels ``using SB'' and ``no SB'' have the same meaning as in~\cref{fig:transv-step1&3}.}
\label{fig:gT-distr-step1&3}
\end{figure}

\cref{fig:gT-distr-step1&3} shows the distribution of $g_T$ values at $Q^2 = 4$ GeV$^2$ for the ``using SB'' and ``no SB'' case. In relaxing the initial constraint on the $N^T_q$ parameters, the $g_T$ distribution broadens. This broadening is due to the changes in the $\delta d_v$ distribution, and mitigates the existing tension between phenomenological calculation and lattice QCD estimates. Indeed, the peak of the ``no SB'' $g_T$ distribution moves toward the range of lattice $g_T$ estimates, and its tail  overlaps with the  lattice QCD range, $0.9 \lesssim g_T \lesssim 1.1$. In this sense, by relaxing the initial request of automatic fulfillment of the Soffer bound, the phenomenological analysis is able to explore portion of the parameter space that are less in tension with $g_T$ estimates on the lattice.

%%%%%%%%%%%%%%%%%%%%%%%%%%%%%%%%%%%%%%%%
%%%%%%%%%%%%%%%%%%%%%%%%%%%%%%%%%%%%%%%%

%%%%%%%%%%%%%%%%%%%%%%%%%%%%%%%%%%%%%%%%
%%%%%%%%%%%%%%%%%%%%%%%%%%%%%%%%%%%%%%%%

A summary of the results for the tensor charges, $\delta u_v$ and $\delta d_v$, and for the isovector tensor charge, $g_T$ calculated at $Q^2 = 4$ GeV$^2$, is presented in Table~\ref{tab:results}.
Expectation values and standard deviations are calculated using~\cref{eq:E} and the square root of~\cref{eq:V}. The quot\-ed errors are at $2\sigma$.

\begin{table}[htp]
\centering
\begin{tabular}{c c c c}
\toprule
 ~& $\delta u_v$ & $\delta d_v$ & $g_T$ \\
\cmidrule{2-4}
~ &  \multicolumn{3}{c}{$Q^2 = 4\GeV^2$} \\
\midrule
using SB & $0.42 \pm 0.09$ & $-0.15 \pm 0.11$ & $0.57 \pm 0.13$ \\
\midrule
no SB & $0.40 \pm 0.09$ & $-0.29 \pm 0.22$ & $0.69 \pm 0.21$ \\
\bottomrule
\end{tabular}
 \caption{Summary of the results at $Q^2 = 4\GeV^2$ for the tensor charges and the isovector tensor charge calculations, under the ``using SB'' and the ``no SB'' hypotheses. Expectation values and standard deviations are calculated using~\cref{eq:E} and the square root of~\cref{eq:V}. The quoted errors are at $2\sigma$.
}
\label{tab:results}
\end{table}

%\begin{table}[htbp]
%\centering
%\begin{tabular}{c c c c}
%\toprule
% ~& ${\delta u_v}$ & ${\delta d_v}$ & ${g_T}$ \\
%\cmidrule{2-4}
%~ &  \multicolumn{3}{c}{$Q^2 = 4$ GeV$^2$, $0.035 %\lesssim x \lesssim 0.29$} \\
%\midrule
%using SB & $0.30 \pm 0.03$ & $-0.16 \pm 0.01$ & %$0.46 \pm 0.03$ \\
%\midrule
%no SB & $0.29 \pm 0.03$ & $-0.23 \pm 0.10$ & $0.52 %\pm 0.10$ \\
%\bottomrule
%\end{tabular}
% \caption{
%Summary of the results at $Q^2 = 4$ GeV$^2$ for the %\emph{truncated} tensor charges and the \emph{truncated} isovector tensor charge, calculated over the experimental $x$-range, $0.035 \lesssim x \lesssim 0.29$, under the ``using SB'' and the ``no SB'' hypotheses. Expectation values and standard deviations are calculated using~\cref{eq:E} and the square root of~\cref{eq:V}. The quoted errors are at $2\sigma$.
%}
%\label{tab:results-trunc}
%\end{table}

A word of caution and some comments are in order. There are in fact some aspects to be stressed, that would help in enlighten the current knowledge on transversity and on tensor charges.

As already mentioned, the covered $x$ range in the phenomenological extractions is quite limited, namely $0.035 \lesssim x \lesssim 0.29$. This means that, when calculating $\delta q$ and $g_T$, most of the computation is given by an extrapolation based on the adopted model and outside this $x$ range. In this respect, loosening some initial constraints can help in reducing the effect of such extrapolation, but also lead to different results and, in turn, different interpretation. Furthermore, we have to stress that lattice calculations are also based on some specific assumptions such as choice of the action, lattice spacing, etc, and that are performed considering matrix elements over the full $x$ range. Therefore, the comparison between phenomenological and lattice results should be done prudently.

We also notice that a similar analysis has been performed by including lattice data on $g_T$ directly into the fit procedure~\cite{Lin:2017stx}. The two transversity parametrizations used here and by Lin et al.~are quite similar, but the fit of Ref.~\cite{Lin:2017stx} was performed with different sets of fit parameters and different choices for the collinear PDFs and FFs. Moreover, in order to impose the SB, we parametrize the transversity proportional to the SB itself, while Ref.~\cite{Lin:2017stx} used a generic $x$-dependent form. Nonetheless, the results presented in Fig.~3 of Ref.~\cite{Lin:2017stx} are compatible with ours. Notice that in Ref.~\cite{Lin:2017stx} only SIDIS data were considered, whereas we use both SIDIS and $e^+e^-$ data. 
%It is natural to expect that all fits would converge to similar results if the same data set is used and the same set of functions is being extracted.

It would be certainly interesting to extend such a kind of study to similar analyses performed in the collinear framework, such as the one by Bacchetta and Radici~\cite{Radici:2018iag}, where independent datasets are used and where, within a different parametrization for transversity, the automatic fulfillment of the Soffer bound is also achieved. In fact, the recent study of Ref.~\cite{Benel:2019mcq} does focus on the influence of the SB on the extraction of transversity. The results of the current analysis are in agreement, within the errors, with those of Refs.~\cite{Radici:2018iag, Benel:2019mcq}. Notice that the contribution of down quark varies the most between different studies, ours and Refs.~\cite{Anselmino:2008sga, Lin:2017stx,Radici:2018iag, Benel:2019mcq}. This is not unexpected: down-quark function is less constrained by the experimental data, the bound is more stringent, and thus one has to expect the larger variation of results depending on the methodology of the fit and the parametrization used. %

\section{\label{sec:conclusions}Conclusions}

In this letter we have studied the role of initial assumptions in phenomenological analyses for transversity function from SIDIS data. The tranversity distributions are usually parametrized in terms of their corresponding Soffer bounds and, upon some choices, the bound is automatically fulfilled throughout the analysis. 

By means of an explicit example, we have shown that, by relaxing the initial assumptions on the parameters that ensure the automatic fulfilment of the bound, we could obtain interesting information on the size of the violation of the Soffer bound observed in current SIDIS data. It turns out that there is no statistically significant violation of such bound. Moreover, loosening the initial choices on the parameters has allowed us to explore the parameter space more accurately and get more reliable estimates on the errors, in particular for the down quark transversity. 

Using then the extracted transversity functions, we have calculated the tensor charges for $u_v$ and $d_v$ quark flavours and, consequently, the isovector tensor charge, $g_T$. Another effect of loosening the initial constraints on the parameters for transversity is on the tensor charge estimates. In fact, the existing tensions observed between phenomenological and lattice QCD estimates of $g_T$ are eased, and the $g_T$-value distribution moves towards the range of lattice QCD estimates. Nonetheless, the discrepancy observed for individual contributions, and in particular for $\delta u_v$, persists. In order to resolve this issue one will need to perform new phenomenological analyses in different approaches, studying the underlying assumptions of both phenomenological analyses and lattice QCD calculations. For instance, suitable choices for the large-$x$ behaviour of $h_1^q(x)$ may help in reducing the aforementioned discrepancies for $\delta u_v$.

Current SIDIS data are, at the moment, not sufficient to constrain the valence transversity functions and, in turn, the tensor charges. The probed $x$ range, ($0.035 \lesssim x \lesssim 0.29$), is still too narrow to avoid the effects of extrapolations made in the integration needed to compute tensor charges and $g_T$. If we calculate truncated values for ``no SB" fit of $\delta u_v$, $\delta d_v$, and $g_T$, we obtain $0.28 \pm 0.06$, $-0.21 \pm 0.17$, and  $0.50 \pm 0.17$ respectively (to be compared with the values in Table~\ref{tab:results}). Thus, around 30\% of the tensor charge value results from an extrapolation to an unexplored region of $x$.

In this respect, new SIDIS data from the future Electron Ion Collider and Jefferson Lab could definitely help in reducing the effect from this model dependence and expand the region of explored values of $x$. Another avenue of constraining transversity is the addition of data from other processes, where transversity is probed into a global fit, such as the left-right asymmetry measured in polarised proton-proton scattering, within the twist-3 collinear formalism~\cite{Kanazawa:2014dca,Gamberg:2017gle} or within a phenomenological TMD framework~\cite{Anselmino:2012rq}.
%Furthermore, given the current situation for the data, comparisons between phenomenological calculation and lattice QCD estimates should be done carefully.

In conclusion, similar educated analyses in different frameworks would certainly be helpful in pinning down the transversity function and, in turn, constraining the tensor charges and the isovector one more reliably. 
%In doing that, hadron physics could become even more helpful for testing non-standard, beyond Standard Model effects.

\section*{Acknowledgments}
We would like to thank F.~Murgia and D.~Pitonyak for helpful discussions and careful reading of the manuscript. This paper was supported in part by the National Science Foundation under Grant No.~PHY-1623454 (A.P., C.F.), the DOE Contract No.~DE-AC05-06OR23177 (A.P.), under which Jefferson Science Associates, LLC operates Jefferson Lab, by the European Union’s Horizon 2020 research and innovation programme under grant agreement No.~824093 (C.F.), and by Regione Autonoma della Sar\-degna (C.F.). C.F.~is thankful to Penn State Berks for hospitality and support for his visit during which part of the project was done.

\bibliographystyle{elsarticle-num}
\biboptions{sort&compress}
\bibliography{h1-SB}

\end{document}